\documentclass[12pt]{iopart}

\usepackage{iopams}

\usepackage{graphicx}
\usepackage{harvard}

\begin{document}

\title[Drop bouncing off inclined flowing soap films]{On angled bounce-off impact of a drop impinging on a flowing soap film}

\author[Basu, Yawar, Concha, Bandi]{Saikat Basu$^1$\footnote[1]{\scriptsize{Corresponding author, currently at the Computational \& Clinical Fluids Research Lab, Department of Otolaryngology, University of North Carolina at Chapel Hill -- School of Medicine, Chapel Hill, North Carolina 27599, USA}}, Ali Yawar$^1$\footnote[2]{\scriptsize{Currently at the Biomechanics and Control Lab, Department of Mechanical Engineering \& Materials Science, Yale University, New Haven, Connecticut 06520, USA}}, Andres Concha$^{2, 3}$ and M M Bandi$^1$}

\address{$^1$Collective Interactions Unit, Okinawa Institute of Science and Technology Graduate University, Onna-son, Okinawa 904-0495, Japan\\
                $^2$Condensed Matter i-Lab, Diagonal las Torres 2640, Building D,  Pe\~{n}alolen, Santiago, Chile\\
	     $^3$School of Engineering and Sciences, Universidad Adolfo Iba\~{n}ez, Diagonal las Torres 2640, Pe\~{n}alolen, Santiago 7941169, Chile}
\ead{saikat$\_$basu@med.unc.edu}
\vspace{10pt}
\begin{indented}
\item[]May 2017
\end{indented}

\begin{abstract}
Small drops impinging angularly on thin flowing soap films frequently demonstrate the rare emergence of bulk elastic effects working in-tandem with the more common-place hydrodynamic interactions. Three collision regimes are observable: (a) drop piercing through the film, (b) it coalescing with the flow, and (c) it bouncing off the film surface. During impact, the drop deforms along with a bulk elastic deformation of the film. For impacts that are close-to-tangential, the bounce-off regime predominates. We outline a reduced order analytical framework assuming a deformable drop and a deformable three-dimensional film, and the idealization invokes a phase-based parametric study. Angular inclination of the film and the ratio of post and pre impact drop sizes entail the phase parameters. We also perform experiments with vertically descending droplets impacting against an inclined soap film, flowing under constant pressure head. Model predicted phase domain for bounce-off compares well to our experimental findings. Additionally, the experiments exhibit momentum transfer to the film in the form of shed vortex dipole, along with propagation of free surface waves. On consulting prior published work, we note that for locomotion of water-walking insects using an impulsive action, the momentum distribution to the shed vortices and waves are both significant, taking up respectively 2/3-rd and 1/3-rd  of the imparted streamwise momentum. In view of the potentially similar impulse actions, this theory is applied to the bounce-off examples in our experiments, and the resultant shed vortex dipole momenta are compared to the momenta computed from particle imaging velocimetry data. The magnitudes reveal identical order ($10^{-7}$ N$\cdot$s), suggesting that the bounce-off regime can be tapped as a simple analogue for interfacial bio-locomotion relying on impulse reactions.
\end{abstract}

%
\vspace{2pc}


\section{Introduction}\label{s:intro}

\begin{figure}
\centering
\includegraphics[width=13cm]{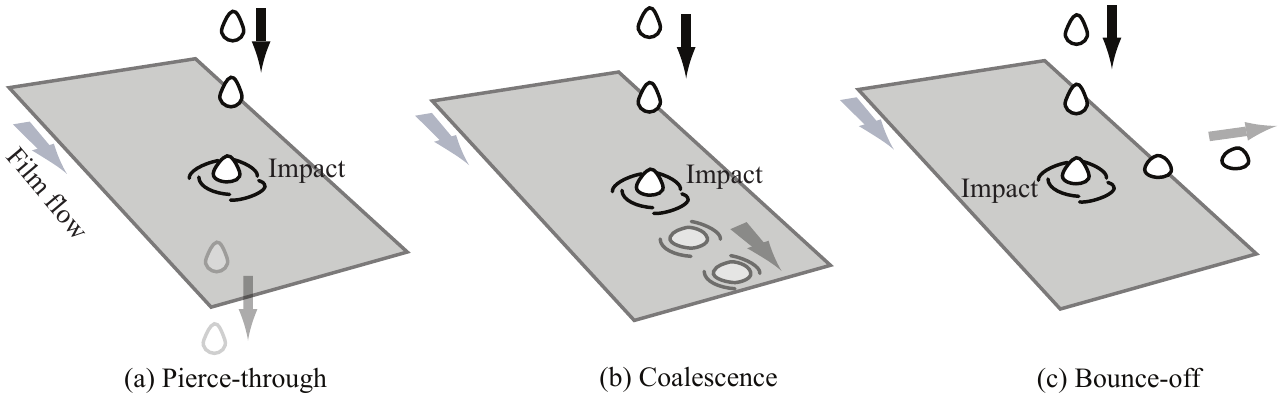} 
\caption{\small{Artistic representations of the three broadly observable collision regimes when drops impinge on a soap film: (a) pierce-through, (b) coalescence, and (c) bounce-off. 
In each panel, the vertical dark arrow represents the drop's pre impact descent direction, the slanted arrow beside the soap film (colored light grey) shows its direction of flow, and the grey slanted arrow adjacent to the post impact drop indicates the corresponding drop motion.}}\label{f:regimes}
\end{figure}

Interactions where the bulk elastic properties of a fluidic system as well as the hydrodynamic effects assume comparable significance are a rarity. Consideration of elastic effects in problems of fluid mechanics is, in most situations, restricted towards accounting for the surface stretching that leads to the interfacial tension. For a soap film, owing to the minuscule thickness, it is often a robust idealization to assume that its dynamic behavior is similar to that of a stretched two-dimensional membrane, thereby implying that considering bulk elasticity would be superfluous. This assumption can be traced back to various articles on soap film dynamics and interactions, see for e.g. works by \citeasnoun{bandi2013pendulum}, \citeasnoun{salkin2016generating}, \citeasnoun{stremler2011mathematical}, \citeasnoun{stremler2014point}, \citeasnoun{couder1989hydrodynamics}. However, to gain a physical understanding of a more complicated interaction, the current work proposes a model system which demonstrates in-tandem effects of both hydrodynamics and the bulk elasticity. It involves very small drops (with diameters approximating 1.20 -- 1.40 mm) impinging into a flowing soap film, wherein the nature of post impact dynamics is affected by the three-dimensionality of the film material. During the impact, the drop shoots into the film surface and deforms it, thereby generating a bulk restitutive reaction. The drop also deforms and spreads out on impact. Salient hydrodynamic effects like shed vortices, owing to the shear layers generated from the relative motion between the gravity-driven drop and the film flow, emerge during the interaction. These interactions can be broadly classified into: (a) the falling energy of the drop is high enough to make it tunnel down through the film (\textit{piercing} regime), (b) the drop hits the film surface and moves downstream with the background stream (\textit{coalescence} regime), and (c) it hits the film surface, moves downstream for a finite time, and then disconnects and bounces off the surface (designated as the \textit{bounce-off} regime). See Figure~\ref{f:regimes} for a schematic representation of these three major regimes of drop impact.

It is important to note that a larger drop size significantly complicates the dynamics leading to a ``pinch-off'' regime wherein a part of the drop connects with the flow and is sheared away while the rest of the drop mass persists to track an inertial trajectory off the film surface. \citeasnoun{chen2006pof} have studied the influence of viscosity on such pinch-off effect. Such interactions, often also referred to as ``partial colescence'' (see, for e.g. \citeasnoun{thoroddsen2000pof}, \citeasnoun{honey2006pre}, \citeasnoun{gilet2007pre}), are however beyond this work's purview.

\begin{figure}
\centering
\includegraphics[width=13cm]{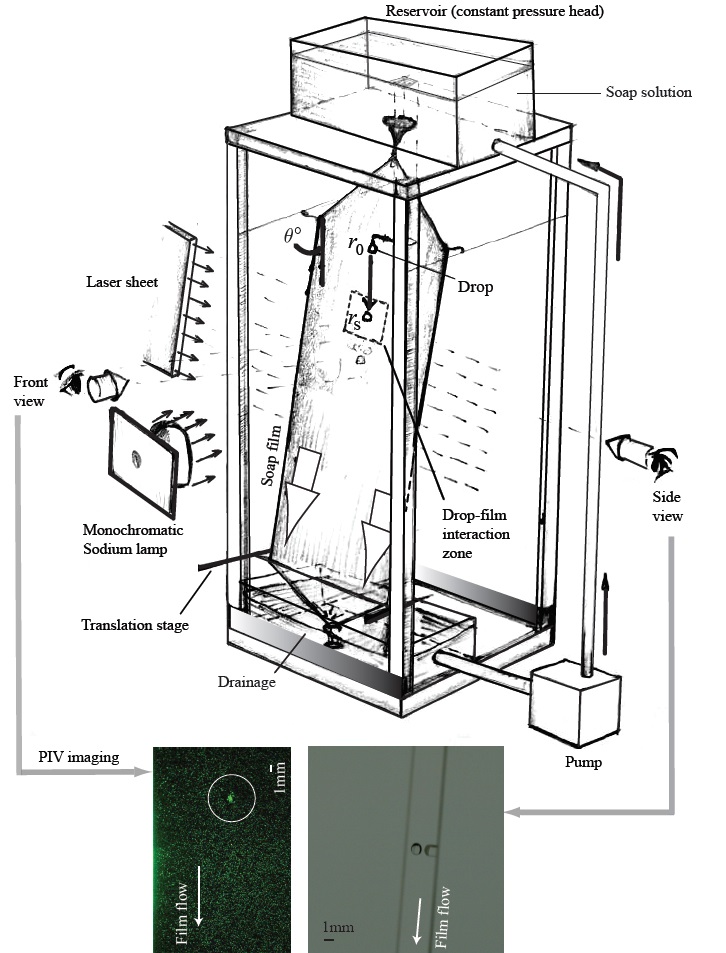} 
\caption{\small{An artistic rendering of the experimental setup. The soap water flows steadily through a wired network under a constant pressure head maintained at the overhead reservoir through a re-circulatory pumping system. Width of the film is 5 cm, and the length is approximately 194 cm. Soap solution concentration is 2.5\% by volume in deionized water. Two cameras were used to record the interaction. Side view: along the plane of the film, which was lit by diffused white light, and front view: perpendicular to the plane of the film lit by monochromatic sodium light with wavelength $\lambda = 532$ nm. For flow field measurements through the particle imaging velocimentry technique, an optical setup produced a laser sheet passing through the film and recorded by the front camera. Inset snapshots at the bottom representatively depict the corresponding camera visualizations. Please note that the sketch is \textit{not-to-scale}.}}\label{f:setup}
\end{figure}

We have performed a series of experiments where the drop hits the film surface at a shallow angle ($<10^{\circ}$), which is the complementary angle subtended between the film ``plane''\footnote[3]{Note that the so-called film ``plane'' is not a truly planar surface. The film has undulating ripples and typically may have a subtle sagging. When we mention the angle subtended by the film's plane with the vertical, it is essentially the angle between the wire, holding the film, and the vertical.} and the vertically downward pre impact trajectory of the drop. The angle has been marked as $\theta$ in Figure~\ref{f:setup} - an artistic schematic of the experimental setup. Bounce-off impact predominates for such close-to-tangential impacts. A quasi two-dimensional projection of the resultant impact is remarkably similar in several aspects to the impulsive interaction observed in water strider locomotion at air-water interfaces, as studied for e.g. by \citeasnoun{hu2003nature}. In this context, see Figure~\ref{f:momentum}(a) and ~\ref{f:momentum}(b) for the congruous features between the two apparently anomalous systems.  For higher angles of impact, the pierce-through regime gains precedence, as pointed out in the work of \citeasnoun{kim2010pre}.

The components of this article can be schematized into: (a)~experimental visualization of drops impacting the film at shallow angles, (b)~developing a reduced order mathematical model to identify the phase space for the observed bounce-off regime, with the parameters coming from angular inclination of the film and the on-impact deformation features of the drop, (c)~comparing the model predictions with experimental observations, thereby gaining an insight into the efficacy of the idealizations, and (d)~identifying the relevance of the impulsive nature and momentum transfer observed in this bounce-off regime with interfacial locomotion of animals living at the water-air interface\footnote[7]{Note that Gerridae, some beetles, are however not classified as ``aquatic''\cite{cheng1985}, despite portraying such air-water interfacial locomotion.}. In this context, for additional study on similar bouncing interactions, see \citeasnoun{courbin2006prl}'s work related to bouncing of solid beads on a stationary elastic membrane. \citeasnoun{jayaratne1964} and more recently \citeasnoun{pan2005asme} have explored the criteria for bouncing versus coalescence for droplets striking the free surface of a fluid bath, with the liquid layer backed off by a solid surface. Chaotic bouncing of a droplet on a soap film has been investigated by \citeasnoun{gilet2009prl}, along with a more intensive treatment in \cite{gilet2009fluid}. More recently, \citeasnoun{gilet2012pof} have looked at droplets bouncing on a wet, inclined rigid surface with a thin coating of highly viscous fluid.

\begin{figure*}
\centering
\includegraphics[width=15.5cm]{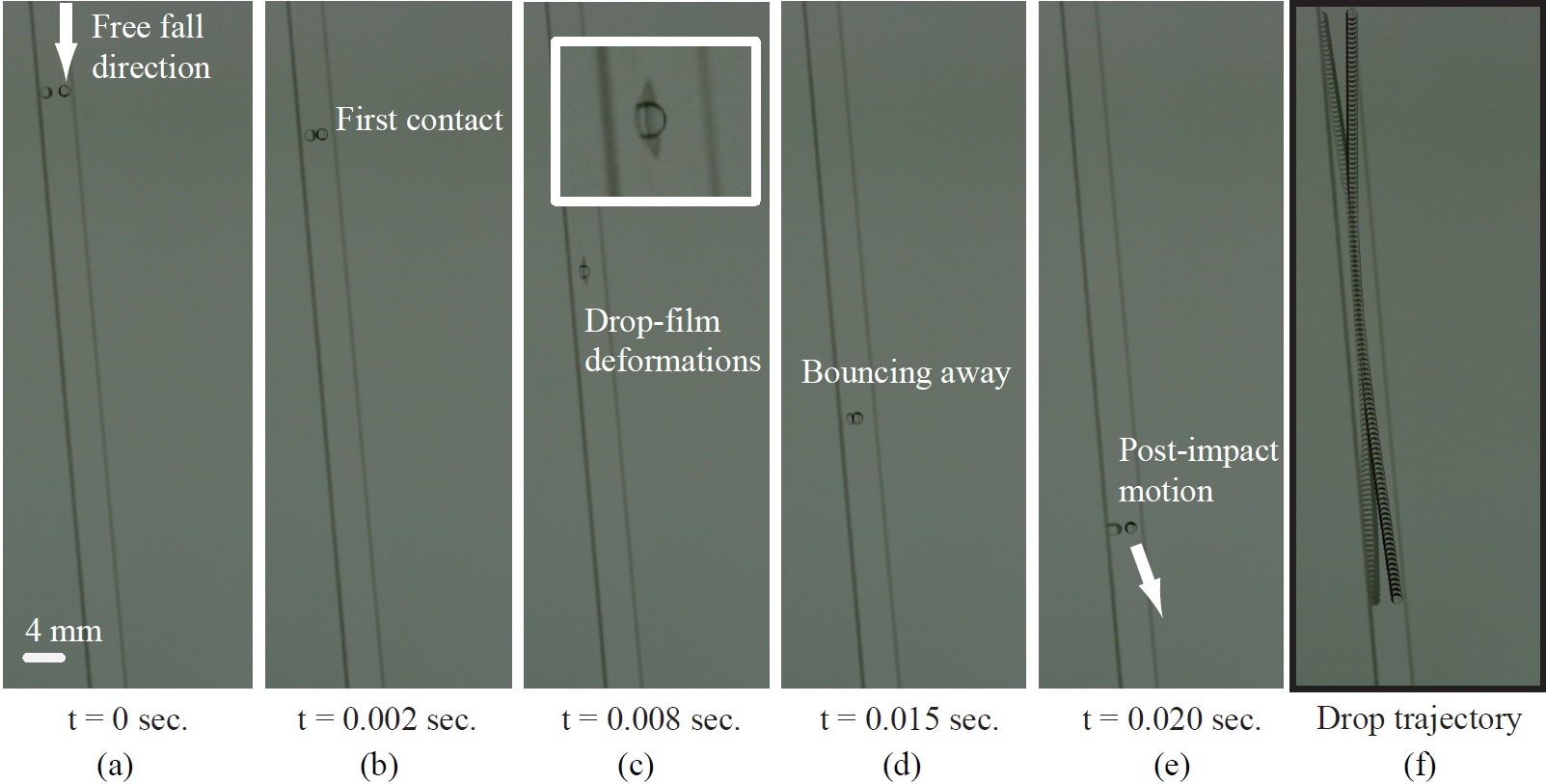} 
\caption{\small{A representative experiment depicting the different stages [(a)--(e)] during the bounce-off regime for a vertically descending droplet impinging on an inclined flowing soap film. Imaging is done with the line of sight being almost in plane with the film, from its lateral side. The different time steps for each of the snapshots are labeled. The droplet deforms the film and is itself maximally deformed about halfway through the contact phase. Panel (c) provides a zoomed-in view of the respective deformations. Panel (f), marked out by the dark boundaries, provides a more comprehensive illustration of the impacting drop trajectory, through superposition of images captured at regular time increments.}}\label{f:impact}
\end{figure*} 


\section{Experimental parameters and observations}\label{s:experiments}

\subsection{Soap film setup}
The soap film setup consists of two parallel nylon fishing lines (of thickness 0.3 mm and length $\sim$ 2.0 m, each), hanging taut with a spanwise separation of 5 cm. On the top, the lines are tied together to a plastic reservoir containing soap solution (Dawn$\textsuperscript{TM}$ dishwashing soap dissolved in deionized water with a concentration of 2.5\% by volume), through a flow control valve. The overhead reservoir has a level sensitive drain, and is constantly refilled with soap solution that drains to the bottom of the fishing lines, where a hanging mass keeps them taut. Four nylon guide strings wrap around the cords to keep them separated. By controlling the flow rate and maintaining a constant pressure head, we ensure a steady flowing soap film.  Interferometry results suggest that the film thickness is roughly in the order of microns. The averaged pre impact drop diameter is approximately $1.4 \pm 0.2$ mm. The observational errors can be traced to the annular thickness of the drop ``halo", owing to the imaged bright peripheries of the drop surface. The film length is 194 cm streamwise and the cross-stream film width is 5 cm, the latter being controlled by the transverse stream separation of the wires. The experimental rig is constructed using commercially available 80/20 pieces. We use fresh soap solution, and the plumbing is flushed with de-ionized water before every trial to keep the system free of debris (like rust particles).

By controlling the horizontal displacement of the bottom end with respect to the top using a translation stage, the film is inclined at 8 different angles ($0.8594^{\circ}$ -- $6.8528^{\circ}$) to the vertical. Above these angles, the film started to sag.  As mentioned already, for higher angles of impact, the puncture of the film leading to the drop tunneling through it becomes a more common occurrence. Figure~\ref{f:impact} demonstrates a representative bounce-off case, imaged from the side with the line of sight being almost in-plane with the film orientation. Later on in this manuscript, the reader will find Figure~\ref{f:dipole} (imaged frontally) which demonstrates the vortex dipole shed by the drop when it is in contact with the background film flow.

\subsection{Drop generation}
A Terumo 33G syringe needle connected to a syringe pump via a hose is used as the source of deionized water droplets. The ejection conditions are so monitored as to ensure zero horizontal velocity of the drop. We can estimate the rotational inertia of the drop as $\frac{2}{5} \,m\, r^2 = \frac{2}{5}(\rho \,V)\,r^2$, which is about $10^{-12}$ kg.$m^2$. Here, $m$ is the mass of the drop, while $\rho$ denotes the material density and $V$ is the volume of the impacting drop. Assuming a minimum droplet velocity of about 1 m/s, the energy is about $10^{-6}$ J. In order to have rotational energy comparable to the overall kinetic energy, the angular velocity of the drop must be about $10^3$ rad/sec, which is unrealistic. Thus, the spin effects can only account for a negligible energy contribution. Flow rate is chosen to ensure that successive droplets are temporally separated in their impact with the film, so as to avoid unwanted wake interactions on the substrate. The needle is mounted on a double axis horizontal-vertical translation stage. Three heights of drop release are selected to achieve three velocities of drop impact. We have 48 bounce-off data-points obtained for these 3 different heights of free descent for the drop, combined with 8 angles of film inclination for each, with 2 trials in each orientation. The imaging protocol includes observing thickness fluctuations (through interferometry) and flow fluctuations (through particle imaging velocimetry). For front and side imaging of the impact, we have used digital high speed cameras (Vision Research Phantom v641), with a typical imaging speed range of 4500--6000 FPS.

\subsection{Diffused light imaging}
The soap film is illuminated normally by a monochromatic sodium light source (with wavelength $\lambda = 532$ nm), the images being recorded at 4500 FPS (front view, refer to Figure~\ref{f:setup}). In plane, a diffused white light tablet illuminates the film (``side image", refer to Figure~\ref{f:setup}), and the imaging is done by another high speed camera at 4500 FPS. The diameter of the nylon cables is used to determine the length scale of the images. Sodium light interference fringes manifest as contrasting dark and bright regions on the soap film, corresponding to thickness fluctuations, enabling visualization of surface waves and vortices. 

Image processing is performed on the collected images in \textsc{matlab} in order to measure pre impact drop radius $r_0$, semi-major axis at maximum deformation of the drop (assuming droplets to be ellipsoids at maximum deformation) $r_s$ on impact, and its pre and post impact velocities $v_0$ and $v_f$ respectively. Measurement accuracy is constrained by the amount of available contrast between the droplet and its background. For spherical droplets, a circular Hough Transform is implemented to measure the radius; but in the deformed drop case, the droplet height is determined by manually counting the pixels between the edges of the droplet defined by the sharpest gradient of the drop image.

\subsection{Particle Imaging Velocimetry}
Hollow glass spheres, having diameter 10 $\mu$m, are suspended in the soap solution and a 532 nm continuous wave Diode Pumped Solid State Laser at 5--10 Watt illumination is used to conduct the particle imaging velocimetry measurements. The beam is passed through a cylindrical lens to generate a laser sheet, followed by collimation optics to pass the laser sheet from one side of the film to the other. The laser light scattered by the hollow glass tracer particles is captured at 6000 FPS (front view, refer to Figure~\ref{f:setup}). This data is post-processed using Particle Imaging Velocimetry algorithms written in-house to measure the two-dimensional velocity field through a rectangular grid-based Eulerian approach. Quantitative estimates of the shed vortex momenta use the length-scale measurements of the vortex dipole and its speed with respect to the soap film flow, as reported by \citeasnoun{bush2006annrev}.


\section{Modeling framework}\label{s:model}

\begin{figure}
\centering
\includegraphics[width=13cm]{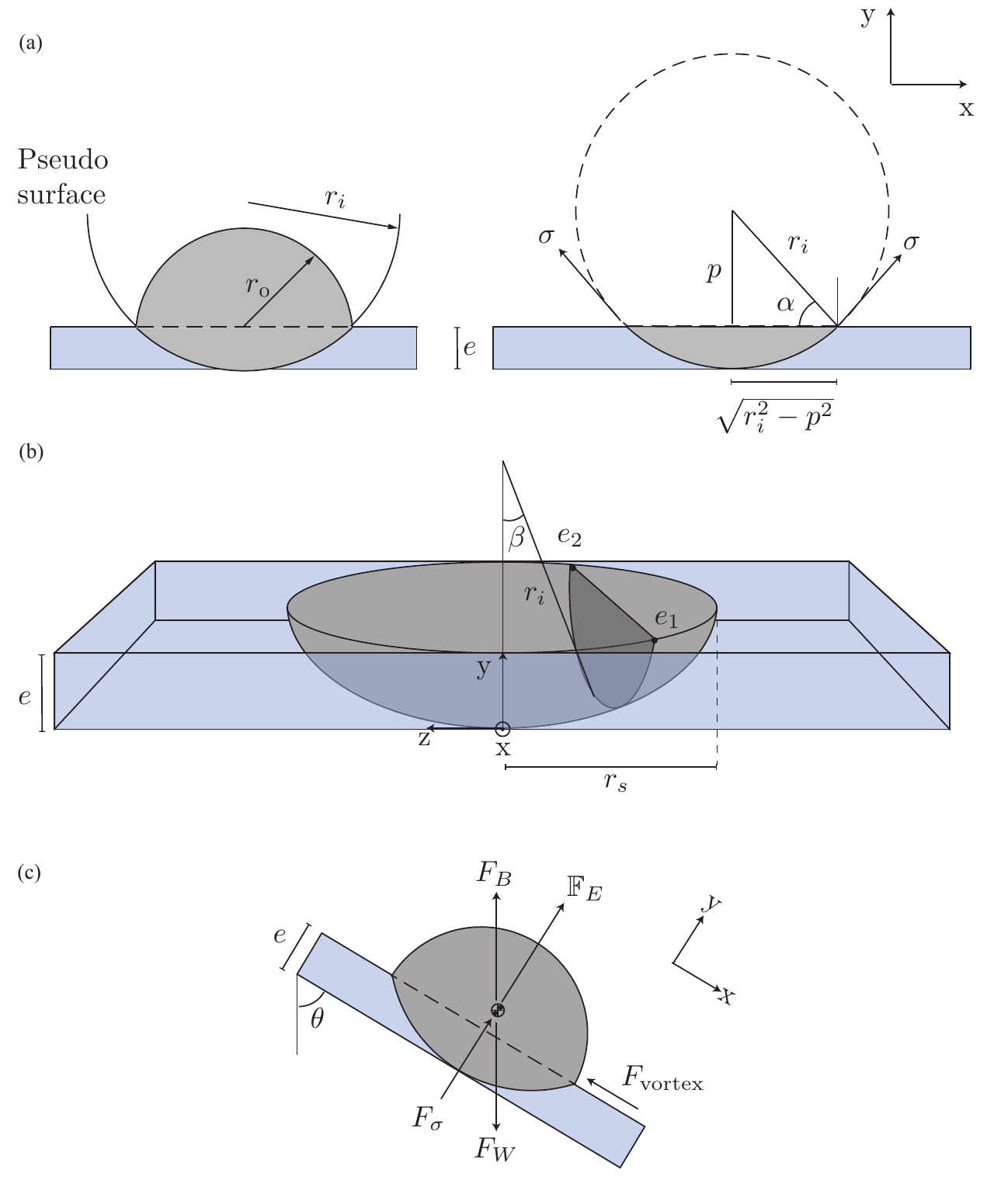} 
\caption{\small{Outlines of the idealized deformed geometries of the drop during its impact on a film of thickness $e$ flowing along the x-axis. The marked ``pseudo surface'' stands for the spherical curvature of the deformed drop volume that has impinged into the film material. See Section~\ref{s:model} for the related details.
(a) Schematic view of the drop deformation on impact. Here, $r_0$ is the radius of curvature of the original pre impact spherical shape of the droplet. $r_i$ is the radius of curvature of the drop volume that impinges into the film thickness.
(b) Perspective view of a impinged droplet sliced along the film plane showing the differential cross section $e_1 - e_2$ of the deformed film surface used to calculate bulk elastic force on the drop. Here $r_s$ is the radius of the spread outline of the deformed drop on the film plane. 
(c) Force balance diagram of the quasi-static deformed droplet in contact with the film inclined at angle $\theta$ to the vertical}}\label{f:calculations}
\end{figure}

\begin{figure*}
\centering
\includegraphics[width=13cm]{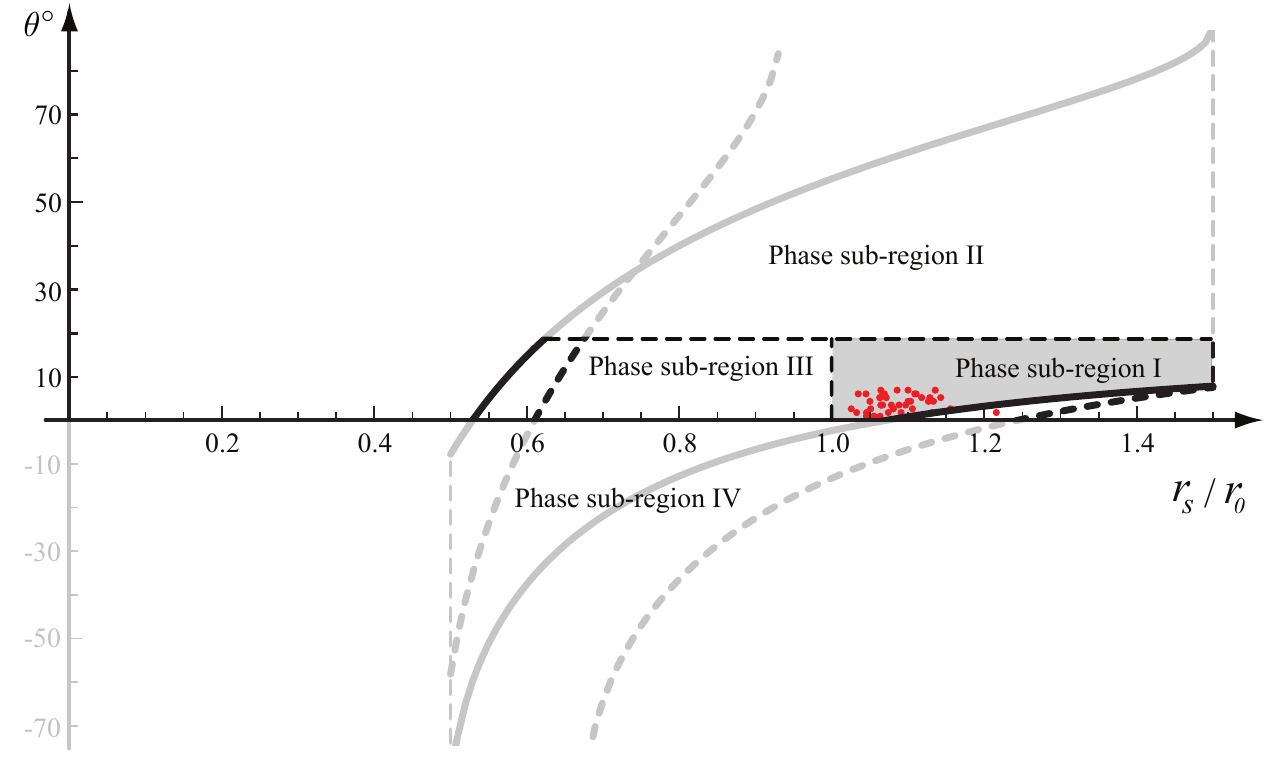} 
\caption{\small{Phase space representation for the bounce-off regime. Vertical axis represents the angle ($\theta$, in degrees) between the pre impact drop trajectory and the inclined soap film. Horizontal axis characterizes the spreading trends of the drop as it hits the film, in terms of the ratio between its post and pre impact radii ($r_s / r_0$). The solid red circles indicate the phase space positions of the 48 experimental data-points. For the input parameter, we use the observed global mean (0.691 mm) as well as a lower estimate (0.6 mm, to account for observational inaccuracies) for the experimental pre impact drop radii ($r_0$). The solid demarcating lines correspond to the observed mean, while the dashed demarcating lines come from assuming the lower estimate. These lines delineate the model region in the phase space, that would correspond to a bounce-off regime. The horizontal axis quantifies the spread of the drop as it hits the film surface. Phase sub-region I (colored grey) marks the parametric space that theoretically corresponds to the bounce-off regime. All but one data-point lie in the predicted model region (intersection of the regions predicted using the two extremal estimates of $r_0$). The sub-regions II--IV represent physically untenable regions for a bounce-off impact. We have used Maragoni elasticity modulus $E_M = 30 \times 10^{-3}$ N/m and surface tension coefficient $\sigma = 73 \times 10^{-3}$ N/m \cite{degennes2013book}. Section~\ref{s:model} discusses the related details.}}\label{f:phaseportrait}
\end{figure*}

To ensure mathematical tractability of the drop-film interaction, we propose a number of idealizations based on the observed experimental features. We assume that the impact does not involve any fluid transfer between the impinging drop and the film substrate. For identifying the parametric region for a specific regime of impact, we conduct a static analysis  of the drop after it has impinged into the film, thereby deforming itself (it spreads out on the film surface) and also the film. It is assumed that the kinetic energy from the droplet descent has been used up to deform the film material and it is at rest for an infinitesimal time. Our analysis captures this stationary point and considers the forces that are acting on the drop. Similar modeling techniques, albeit for different problems of drop impacts have been adopted before, for example by \citeasnoun{chappelear1961models}, who had studied vertical drop impacts and entrance in a liquid bath. Another simplifying assumption entails that at the point of maximum deformation, the deformed tip of the impacting drop just touches the lower surface of the film substrate. Hence, the depth of the peripheral tip of the impinged volume of the drop from the original plane of the film flow is constrained by the thickness of the film. Our experimental design does not allow for an accurate estimation for the film thickness, and hence for the model prediction, we have used published values (1.5 $\mu$m; see for e.g. \citeasnoun{schnipper2009vortex}) for similar setups.

To capture the parameters necessary for the bounce-off regime, the resultant force component perpendicular to the film plane (on the side of the pre impact drop location) should exceed zero. This generates the force inequilibrium criterion: 
\begin{equation}\label{e:phase}
 \mathbb{F}\left(r_s,\theta\right) = F_\sigma (r_s) + \mathbb{F}_E (r_s)+\left[F_B (r_s) - F_W\right]\sin\theta > 0,
\end{equation}
where $\mathbb{F}$ is the resultant force on the impacting drop in the normal direction to the film flow, $F_\sigma$ is the interfacial tensile force component perpendicular to the stream, $\mathbb{F}_E$ is the bulk elastic force component (assuming linearized elastic deformation transverse to the film flow) normal to the film surface and flow direction, $F_B$ is the buoyant force, $F_W$ is the drop weight, and $r_s$ is the spread radius of the intersectional periphery of the drop on film plane. As a valid simplification for this treatment which primarily characterizes drops impacting off the film, the other possible forces acting on the drop along the film flow plane (like drag forces from the shed vortices, apart from the other streamwise force components from above) are not incorporated in the criterion (i.e. inequality~\ref{e:phase}).

Figure~\ref{f:calculations} presents a schematic of the model geometry that facilitates the $(r_s,\theta)$ functional reduction of the different force contributions. Let $r_0$ be the initial radius of the falling drop while considering it to be spherical. As it touches the film surface, the spread-out volume of the drop (impinged into the film thereby deforming it as well) is assumed to have $r_i$ as its radius of curvature (see Figure~\ref{f:calculations}(a)). The remaining volume (which is off the film surface) is assumed to maintain its original radius of curvature ($r_0$). Also, $e$ represents the thickness of the soap film.

Using the above definitions, it can be shown that
\begin{equation}\label{ri}
r_i = \frac{{r_s}^2+e^2}{2e}.
\end{equation}
Let $\rho_1$ be the density of the droplet fluid and $\rho_2$ be the density of the film material. Note that in our experiments with drops constituted from deionized water and the film solution having a soap concentration of 2.5\% by volume, the two densities can be considered to be approximately same and equal to 1000 kg/m$^3$. However, similar experiments can also be carried out using a drop made out of a liquid metal in order to drastically change the dependence on liquid densities (for e.g. \citeasnoun{khoshmanesh2017rsc} have used gallium-based liquid metal for microfluidic experiments). So, for generalization, we keep the two density terms distinct in the mathematical formulations. With $g$ being the gravitational acceleration, the buoyant force on the impacting drop, owing to the displaced film material, can be expressed as
\begin{equation}\label{fb}
F_B= \pi \left[\frac{2}{3}r_i - (r_i - e)\left(1-\frac{(r_i-e)^2}{3{r_i}^2}\right)\right]\rho_2 {r_i}^2 g.
\end{equation}
Thus, using Equation~\ref{ri} in Equation~\ref{fb}, $F_B$ is reducible to a function of $r_s$.

With the model exclusively addressing the bounce-off regime, we only consider the surface tension force that is normal to the film plane. It can be mathematized as
\begin{equation}\label{fsigma}
F_\sigma = 2 \pi \sigma e\left(2-\frac{e}{r_i} \right),
\end{equation}
where $\sigma$ is the tensile force per unit length. Also, the weight of the drop is simply $F_W = \frac{4}{3}\pi\rho_1 g {r_0}^3$.

For the linearized bulk elastic effects of the deformed film material, we just consider the deformation transverse to the film plane, so as to the focus on the bounce-off impact. These restitutive forces on the drop are integrated over the entire deformed contour, to obtain 
\begin{equation}
\mathbb{F}_E = 4\, E_M \int\limits_{0}^{\cos^{-1}(1-\frac{e}{r_i})} \int\limits_{0}^{\mathcal{G}}\left(e-\left[\sqrt{r_i^2-x^2}+r_i(2-\cos{\beta})\right]\right) dx\,d\beta,
\end{equation}
where
\begin{equation}
\mathcal{G} = \sqrt{2er_i(2-\cos\beta)-r_i^2(\cos^2\beta-4\cos\beta+3)-e^2}.
\end{equation}
Here the angular parameter $\beta$ tracks out the spherical annulus of the impinged drop contour from the center of curvature of the impinged volume (see Figure~\ref{f:calculations}(b)). Note that since the impact and the resultant deformation occur over a short time scale in a thin substrate; therefore it is contemplated feasible and sufficient to just consider Marangoni elasticity ($E_M$). See recent contribution by \citeasnoun{kim2016marangoni} on measurements of Marangoni elasticity in flowing soap films. 

\subsection{Phase space representation}
Angular inclination of the soap film and the non-dimensional spreading parameter (comprising the ratio of post and pre impact radii) of the drop chart out the phase space domain in which we explore the post impact drop behavior. Figure~\ref{f:phaseportrait} shows the phase portrait representation depicting the physical parametric bounds that lead to a bounce-off. The region enclosed by the curves is determined from the force criterion on $ \mathbb{F} (r_s, \theta)$, vide inequality~\ref{e:phase}. Ratio $r_s / r_0$ quantifies the spreading of the drop as it impinges into the film surface. 

For phase sub-region II, the higher angle of inclination would lead to a sagging of the film, which can induce incoming contaminating flows from the peripheries on to the test section for drop-film interaction. Such complexities are beyond the scope of our model framework. In phase sub-region III, $r_s<r_0$ defies the experimental observation that the small drops mostly spread out on impact. Phase sub-region IV consists of negative angles of inclination and is merely a ``mathematical construct" owing to the dependence of $\mathbb{F}$ on $\theta$, whereby the phase space contour limits in Figure~\ref{f:phaseportrait} are based on the trigonometric restriction $\sin\theta \in \left[-1, 1\right]$. 

Phase sub-region I (colored grey in Figure~\ref{f:phaseportrait}) represents the predicted model parameter space for the observed bounce-off regime. Its vertical extent is marked by the dashed line at $20^{\circ}$ inclination. However, it is an open question as to at what angle the film sagging becomes appreciable (resulting in contamination of the impact region from peripheral flows) and this demarcation is just a very conservative estimate. The horizontal extent of the region is determined based on the observation that the drop always deforms and spreads out (i.e. $r_s > r_0$). Comparison with experiments turns out satisfactory as all the experimental data-points, except one, could be located in the predicted phase domain.


\section{\label{momentum}Momentum transfer and natural world analogues}

\begin{figure*}
\centering
\includegraphics[width=13cm]{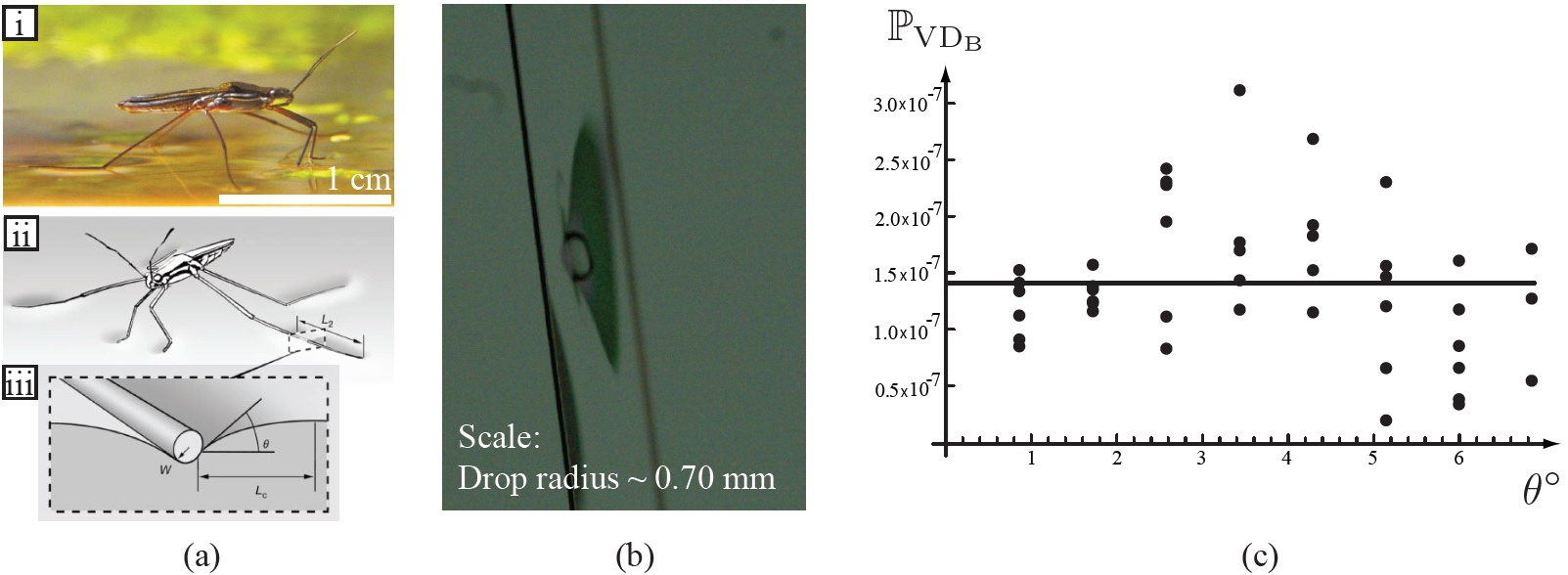} 
\caption{\small{(a) Panels (i), (ii), and (iii) in (a) have been reproduced, with copyrights permission, from \citeasnoun{hu2003nature}. Panel (i) shows a natural water strider (\textit{Gerris remigis}) at the air-water interface. Panels (ii) and (iii) present a cartoon of the interfacial distortion when the strider sits static at the air-water interface, but is about to ``jump". White scale bar (see panel (i)) is 1 cm. (b) A representative bounce-off regime as observed in our drop-film impact experiments. The drop deforms and spreads out, as it impinges in and deforms the soap film. (c) Experimental dipole momenta computed from  B\"{u}hler's theory (see \citeasnoun{buhler2007jfm}), using the observed mean pre impact drop radius of 0.691 mm. The average momentum comes out as $1.421 \times 10^{-7}$ N$\cdot$s (shown by the black horizontal line). Momenta values retain the same order even if we consider the lower estimate for $r_0$, which is $0.60$ mm. The averaged vortex dipole momenta then comes out as $0.930 \times 10^{-7}$ N$\cdot$s. The pre and post impact velocities of the drop are used to measure the momentum transfer to the film during this impulse action, subsequently from which the momenta proportions, shared between the shed vortices and the capillary waves, are computed. $\theta$ (in degrees) on the horizontal axis marks the experimental angles subtended by the soap film and the pre impact drop trajectory. Along the vertical axis, we plot $\mathbb{P}_{\mathrm{VD}_\mathrm{B}}$ which represents the shed vortex dipole momenta in the experimental impacts, computed according to B\"{u}hler's theory.}}\label{f:momentum}
\end{figure*} 

\begin{figure}
\centering
\includegraphics[width=13cm]{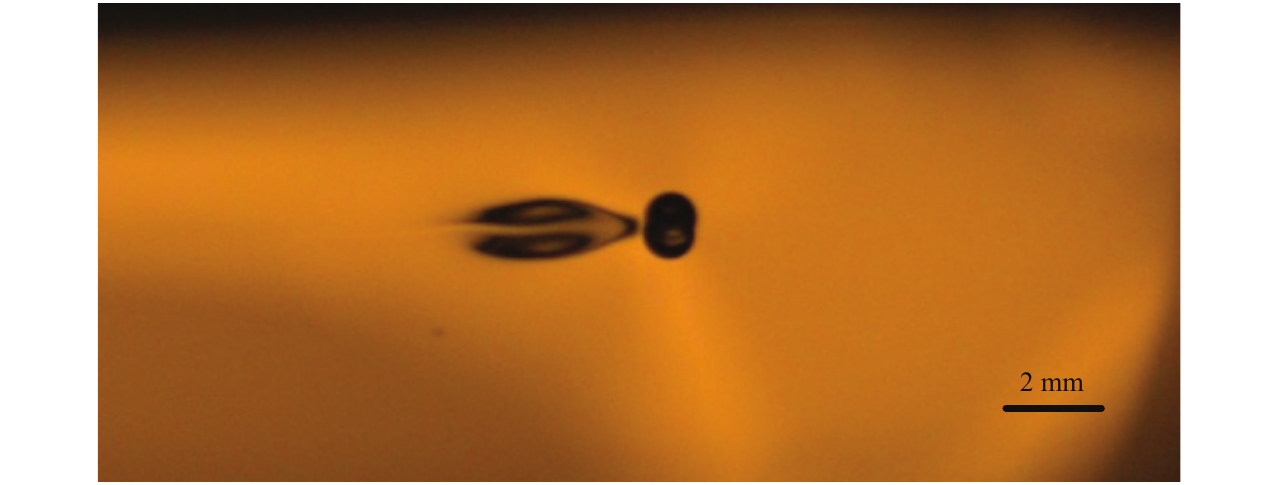} 
\caption{\small{The drop acts like a bluff body on the film flow during the duration of impact (owing to the relative velocities between the two) and sheds a vortex dipole (along with the generation of capillary waves). The above image is captured using a monochromatic sodium lamp as the source of light. The thickness fluctuations assist in determining the vortex cores. Particle Imaging Velocimentry data for similar experiments (using planar laser beam as the light source) give us the velocity field and the accurate dipole momenta can be calculated by tracking the dynamic vortex cores.}}\label{f:dipole}
\end{figure}

\noindent A quasi-static transverse-stream projection of the bounce-off regime is remarkably similar (see Figures~\ref{f:momentum}(a) and \ref{f:momentum}(b)), at least qualitatively, to the impulsive interaction observed in water strider locomotion at air-water interfaces. Figure~\ref{f:momentum}(a), reproduced with permission from  \citeasnoun{hu2003nature}, depicts a schematic of the interfacial interaction when a water-strider jumps off the air-water interface. Like the insect, the drop acts like a bluff body on the interface and generates waves and vortices owing to the relative motion with respect to the background soap film stream. To explore the dynamical connection between these two unrelated interactions, we refer to the theory on impulsive fluid forcing of water walking animals, proposed by \citeasnoun{buhler2007jfm}. The analysis assumes a liquid bath of infinite depth and looks at the interaction at its interface. As expounded in the findings, the insect pushes against the water surface transferring momentum to the bath, and it itself jumps off the surface owing to the impulsive reaction. For the horizontal momentum transferred to the bath, 1/3-rd of it is taken up by the generated free surface waves, and 2/3-rd is accounted for by the shed vortices. We apply the same theory in the current system to estimate the momentum carried by the shed vortex dipole (see Figure~\ref{f:dipole}). The pre impact and post impact velocities of the drop, as measured from the experiments reveal that the drop gains in momentum as it bounces off. We assume that an impulsive momentum of equal magnitude is transmitted to the film. Subsequently, 2/3-rd of the imparted horizontal momentum is calculated to theorize the momenta ($\mathbb{P}_{\mathrm{VD}}$) carried by the shed vortex dipole, as hereunder:
\begin{equation}\label{eom:momentum}
\mathbb{P}_{\mathrm{VD}} = \frac{2m}{3}\left(v_{f_y} \cos\theta + v_{f_x} \sin\theta- v_{0_y} \cos\theta\right),
\end{equation}
where $m$ is the mass of the drop, $v_{0_y}$ is the vertically downward component of the pre impact drop velocity (the horizontal component is monitored to be approximately zero), and  $v_{f_x}$ and  $v_{f_y}$ are respectively the horizontal (positive to the global right, with reference to Figure~\ref{f:calculations}) and vertical (positive downward) components of the post impact drop velocity, just after it has disconnected itself from the film surface. The mean order of the dipole momenta values comes out as $10^{-7}$ N$\cdot$s.

\begin{table}[t]
\caption{Shed vortex dipole momenta for some typical pre impact drop sizes. The values have been derived based on B\"{u}hler's theory (see \citeasnoun{buhler2007jfm}) using the measured pre and post impact drop velocities.} \label{table1} 
\begin{center}
    \begin{tabular}{   l p{6cm} }
    \hline
  Pre impact drop radius ($r_0$) & Shed vortex dipole momentum \\ \hline
0.60 mm &  $0.9303\times10^{-7}$ N$\cdot$s \\ 
  0.65 mm &  $1.1828\times10^{-7}$ N$\cdot$s \\ 
  0.691 mm &  $1.4210\times10^{-7}$ N$\cdot$s \\ 
     0.70 mm &  $1.4733\times10^{-7}$ N$\cdot$s \\ 
    \hline
    \end{tabular}
\end{center}
\end{table}

The estimates from B\"{u}hler's theory are compared to the dipole momenta calculated from particle imaging velocimetry data, through extracting the positional evolution of the coherent vortex structures. These latter values of course provide more accurate quantification for the mometum transfer during the impact. Based on these data, the mean momentum carried by the shed vortex dipole is  $7.425\times10^{-7}$ N$\cdot$s. As a caveat, it should be however noted that this estimate leaves out the data-points for the largest impact angle that approaches break-down of the model idealization (for e.g. the idealization does not consider  any sagging). If we include those numbers, the mean dipole momentum shoots up to $18.191\times10^{-7}$ N$\cdot$s. So, while the order overall matches with the global mean estimate ($1.252\times10^{-7}$ N$\cdot$s. averaged from Table~\ref{table1}) using B\"{u}hler's analysis; the experimental dipole momentum is still higher. The probable reasons can be attributed to the following:
\begin{itemize}
\item Soap film, owing to being stretched taut over the wired network in the experimental setup, manifests a distinct trampoline effect, as excellently discussed by \citeasnoun{gilet2009fluid} in their seminal work. This provides an additional push on the bouncing drop thereby pronouncing the recoil momentum. 
\item The bulk elastic deformation of the film material leads to additional restitutive forces. This effect is absent in static liquid bath of infinite depth, assumed in B\"{u}hler's treatise.
\end{itemize}


\section{Conclusions}

To summarize, we have presented a reduced order model (see Section~\ref{s:model}) to characterize drop impact on inclined flowing soap films, with the core focus being on identifying the parametric region that results in bounce-off dynamics. Angular inclination of the soap film and the on-impact deformation features of the drop quantified by the ratio of its post impact (spread out) and pre impact radii constitute the phase domain parameters. The predicted model phase space fits well with a broad population of experimental observations (Section~\ref{s:experiments} charts out the experimental methods). Figure~\ref{f:phaseportrait} shows the corresponding phase portrait, with the model phase zone for bounce-off drop impact labeled as phase sub-region I. All experimental data-points, except one, belong to this predicted model zone in the phase domain.

Our interest in the problem roots back to the rarity of interactions which evidence in-tandem roles of bulk elasticity and hydrodynamics in a fluidic system, as seen here. Interestingly in this context, the drop-film impact  demonstrates remarkable similarity to the interfacial dynamics observed at the air-water boundary during the impulsive locomotion of water-walking insects, with perhaps the most significant distinguishing feature  between the two phenomena being the varying roles of system elasticity, in lieu of the recoil momentum exerted by the fluid bath on the insect and on the drop. However, quite strikingly the vortex dipole momenta imparted to the fluid substrate reveal the same order ($10^{-7}$ N$\cdot$s) in both the systems. This unexpected congruity suggests that this impulsive bounce-off regime for drops impinging on stretched films may indeed bear relevance as a systemic analogue for investigating the impulse-based locomotion of many aquatic insects living at the air-water interface, like water striders.


\ack Reported findings are based on research conducted at the Okinawa Institute of Science and Technology Graduate University, Japan; under institutional financial support for SB and AY. Additionally, AC was partially supported by Fondecyt 11130075.  


\medskip

\section*{References}
\bibliographystyle{dcu}
\bibliography{fdr2017ref}

\end{document}